# Report on the software "SemanticModellingFramework"

## Publication type:

Report Series

## Authors:

Scalas Andreas

## Source:

https://github.com/andreasscalas/SemanticModellingFramework.git

## Keywords:

Parametric template, Feature-based modelling, Knowledge formalisation, Archaeological reconstruction

## 1.     Introduction

This piece of software is the result of the research work made during a PhD, together with an already published thesis [1], and, for the *fragment fitting* part (see section 3.4), of the participation in the Horizon 2020 GRAVITATE project "Geometric Reconstruction And noVel semantIc reunificaTion of culturAl heriTage objEcts", Grant Agreement n. 665155.

Subject of the PhD research was the definition and implementation of a framework for allowing a semantically-guided modelling of shapes belonging to the same class of homogeneous objects (i.e., sharing specific properties). This is obtained through the formalisation of knowledge regarding the class of objects: indeed, in several contexts, some *domain experts* have deep knowledge about the properties of the objects in the class, indicating characteristics that are common to all of them and without which the objects could not belong to the class.

To exploit such knowledge, our approach has been to first formalise this knowledge in a machine-readable form and then to try and *link* the defined semantics to objects or even some of their parts. This is done following the concept of *part-based annotation* [2], which consists in the association of information (of any kind) to segments of a virtual object, allowing queries to be made to these portions of geometry.

The annotation is a handy method to *build a bridge* between the semantics of the homogeneous classes, which are typically defined as plain textual description of the objects which belong to the class, and the shape of such objects, which in turn encode other useful geometric properties that can be extracted for enriching the knowledge base through *shape analysis*.

Once the knowledge have been formalised and linked it to the geometry of objects in the class, it is possible to define an overall semantised geometry, called *parametric template*, to represent the class and with a variety of possible applications (e.g., classification of shapes based on the comparison with the template, constrained deformation of the template to obtain new shapes in the class, ecc.). The overall geometry can either be selected from one object in the class (archetype) or rather computed as a *statistical model*, encoding all the geometric variability of the shapes in the class.

The remaining of this report is organised as follows: section 2 formalises some terms that are critical for the description of this piece of software; section 3 explains the fuctionalities provided in it; section 4 describes the employed and designed file formats and, finally, section 5 presents an overview of the employed technologies.

## 2.     Definitions

Here some terms that are critical for the presentation of the software are formalised.

- **Mesh**: an approximation of an object's shape, made by a set of vertices, representing spatial coordinates as nodes of a graph, and a set of polygons defining the topology of the graph, i.e., the links between nodes. In this report, meshes are assumed to be *manifold* (see [3] for further info) and that the polygons composing the mesh are only triangles. An explicit reference will be made to *edges*, i.e., the segments linking triangles' vertices;

- **Cage**: is a coarser mesh enclosing the reference shape, typically used for collision detection and meshes deformation (for further details see [4]);
- **Generalized Barycentric Coordinates (GBC)**: GBC are a mathematical tool to describe (better, interpolate) the value of a function in a certain position in space given the values of the function in some *control points*. Several extensions of the concept have been defined (e.g., *Mean Value Coordinates* (MVC), *Green Coordinates* (GC), etc. - see [4] for details);
- **Annotation**: in this work, an annotation is the association of some information to portions of the geometry of a shape. together with a list of properties of the annotation. A *portion of geometry* can either be a (possibly disconnected) region of the surface representing the shape (i.e., a set of triangles of a mesh), a set of feature lines on the surface (i.e., a list of edges of a mesh) or some points of interest over the shape (i.e., vertices of a mesh). The properties (or *attributes*) are quantitative or qualitative characteristics defining some aspect of that specific part (e.g., reference system, site of excavation, notes, etc.).
- **Relationship**: relationships between annotations are of different nature and of different magnitude (i.e, can be between 2 or more annotations), but in general are used to define some common properties among parts (e.g., same symmetry axis, same height), proportions (e.g., the width of the two eyes is similar) or structural configurations (e.g., the head is above the neck, two parts are adjacent, etc.). Finally, hierarchical relationships are used to organize annotations, starting from the whole object and going deeper following the containment between annotations. Another interesting properties of relationships is their *direction*: indeed, not all the relationships are reciprocal, but some may refer to just some of the annotations (e.g., a stylist may design a hat so that it is proportional to the size of the head, but the head is not affected by this relationship).
- **Annotation transfer:** the transfer of annotations is a technique for preserving its geometric selection at the change of resolution of a mesh. This allows to adapt to the requirements of different applications (computational complexity) without losing the annotation work done so far. Details are given in [5].
- **Fragment:** in this report, we call a fragment any secondary mesh that is loaded in the framework and that can be semantically enriched and automatically oriented and posed on the template, eventually deforming the template to fit its shape (see section 3.4). This terminology is due to the main goal of the fragment fitting procedure, which is studied for supporting archaeologists and curators in the reassembly of fragmented cultural assets (see [1]).

## 3. Functionalities

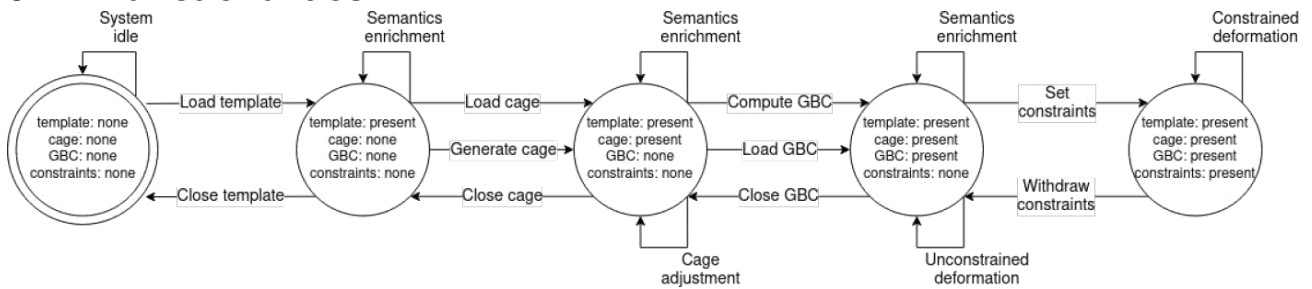

*Figure 1: overview of the system functionalities depicted as states, linked by arrows representing operations. Notice that the "Semantic enrichment" arrow groups the manual annotation, the loading of pre-defined annotations and the extraction of new knowledge through shape analysis techniques.*

The current version of the software, available on GitHub [6], provides all the features presented in the thesis [1]. An overview of the operational workflow and the main functionalities is depicted in Figure 1. After launch, the system starts in the initial state, where no entity is loaded in the application environment yet. After loading a (geometric) template, the user can annotate it. Both a manual annotation functionality and an annotation load function are provided. Support is given to manual annotation providing a set of selection tools (lasso selection of closed surface patches, selection of points or triangles behind a user-drawn rectangle - either applied to visible points/triangles only or to all the points/triangles behind the rectangle - and selection of edges connecting user-picked points) and tagging, with a textual label and a colour. An annotated shape can be saved in an output file (the designed annotation file format is described in section 4) and loaded in future sessions.

Furthermore, annotations can be enriched by attributes. Quantitative attributes can be computed automatically by shape analysis tools. In this software, three measuring tools are implemented, namely the ruler, the tape and the bounding measure. The ruler and tape provide Euclidean and approximate geodesic distance between successively picked points, whereas the bounding measure provides the distance between two clipping planes. Additionally, slicing along a direction and slice analysis functions are available as well as slice clustering methods (see [7]) and OBB extraction. Semantics can be enriched in every state of the system. Then, the cage can also be loaded in the system or generated by a provided simplification and offsetting function. The characteristics of the cage impact the quality of the resulting deformation and the time performance. Therefore, the user may want to select a proper cage resolution and is also allowed to edit cage vertices manually (e.g., to avoid self-intersections or to follow the template shape more closely).

Once the cage is established, the GBC that represent the connection between the cage and the template must be computed. The system implements the automatic generation of the Green Coordinates and the Mean Value coordinates (see [4]). Barycentric coordinates can be saved and loaded in future sessions.

Cage and barycentric coordinates allow at this stage to perform unconstrained cage-based deformation.

This can be done by selecting the cage vertices (or control points) to be manipulated and then translating them and/or rotating them around their barycenter if the selected control points are more than 1. Combinations of these manipulations allow to define any kind of cage configuration. The deformation is propagated accordingly at a fixed frame-rate.

In order to achieve semantics-aware deformation, the user has now to define semantic constraints on annotation attributes and relations. The implemented semantic constraints can be set through the GUI: the user selects the annotations involved and the constraint to be applied with corresponding parameters, including an "importance" factor (weight), to set a priority in case of multiple conflicting constraints. This factor is used as weight into the ShapeOp minimisation for the corresponding geometric constraints and can be used, together with the residual of the cost function of the corresponding constraints, as an indicator for accepting or rejecting a certain deformation.

Finally, the system provides the possibility to select, automatically orient and position *fragments* on the parametric template and then deform the template to try and fit the shape of compatible annotations on the fragments.

This section provides a presentation of the developed system GUI, which offers all the presented functionalities. The proposed GUI is divided in 3 windows, each proposing functionalities that regards a specific task:

- Main window: is the first window shown to the user; here he can load/save a number of meshes (1 main mesh – the parametric template – and any number of other meshes called *fragments*) and, in general, perform any kind of geometry-related tasks, e.g., shape analysis. The main window offers the possibility to load another kind of mesh, a *cage*, that allows to deform the shape of the parametric template;
- Annotation window: in this window, the user can perform all the semantics-related tasks, such as annotation of parts of a shape and definition of relationships between already defined annotations. Finally, the user can *constrain* the already defined relationships, in such a way that they are kept even after shape deformations;
- Relationships window: here the system displays the defined relationships among annotations as a graph, with all the associated info. Moreover, the user can define new relationships and, eventually, constrain the parametric template for deformation.

Each of the above windows will be detailed in a following sub-section, while the last sub-section will be dedicated to the presentation of the fragments *fitting procedure*.

## 3.1 Main window

When the software is run, the main window appears, with a menu bar, a toolbar, a sidebar and a central canvas, initially representing the default shape of a tetrahedron, where objects will be visualised (see Figure 2). The menu bar allows to perform the most common operations (see Figure 3):

- File menu: provides the load, save and close operations for the various entities. The system handles two kinds of entities, namely, barycentric coordinates and 3D models (the template, the cage or other generic meshes representing specific objects in the class or object fragments, which the user wants to investigate);

- Edit menu: includes
    - undo/redo commands;
    - choice and computation of the GBC (actually, the list of generalisations includes MVC and GC, although the selection of the GC generalisation precludes the possibility to use the constrained deformation environment – it can be used for pure geometric deformations);
    - generation of a cage (presently with a simple resample-and-offset approach based on two filters provided in MeshLab [7]: the "Uniform Mesh Resampling", which allows to offset the resulting mesh with respect to the original one (the default is 55% offset with check on "Clean Vertices"), and the "Simplification: Quadric Edge Collapse Decimation" with "Target number of faces" depending on the complexity of the shape and checking on "Preserve Boundary of the mesh" (weight 1), "Preserve Normal" and "Preserve Topology").
- View menu: allows to change the colour of the geometric template and of the cage and to change the visualisation modality: presently the system includes points, edges and surface visualisation, which can be combined as preferred by the user. The default is only surface visualisation for both template and generic meshes and only wireframe and points for the cage.

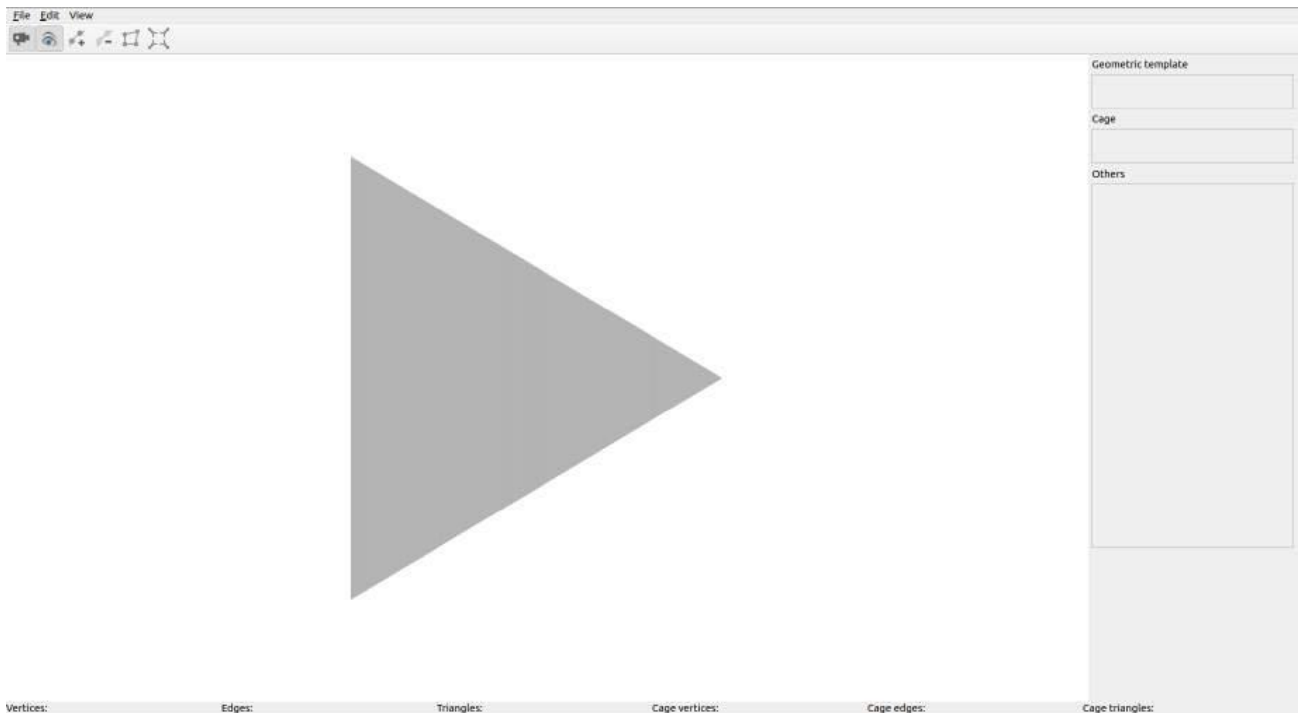

*Figure 2: the main window*

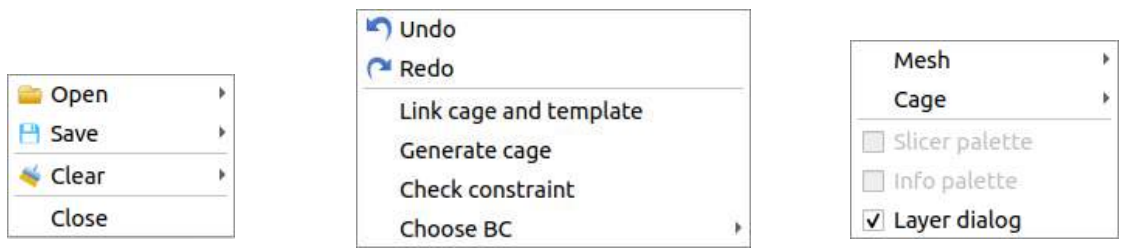

*Figure 3: from left to right: File, Edit and View menu*

The sidebar provides three different views, organised according to the entities involved:

- Layer view: displays the list of meshes actually in use, separated into "Geometric template", "Cage" and "Others": only one geometric template and cage at a time are allowed, while there is no limitation (at least in theory, of course it is restrained by the memory availability) for the other meshes.

- Slice view: provides several shape analysis tools based on the slicing paradigm; these apply to the template model (including the implementation of the method in [8]);
- Constraints view: displays information about the high-level constraint defined by the domain expert.

In the Layer view, the user can interact with each mesh by clicking with the right button of the mouse on it. The interaction includes the possibility to change the visualisation options (as in the View menu) as well as other entity-specific options (see Figure 4):

- Cage: allows to close the mesh (e.g., the user wants to use a finer or a coarser cage);
- Geometric template: in addition to the cage options, it exhibits the "Edit annotation" functionality, which allows to open the Annotation Window (see next sub-section), and the "Show annotation" checkbox, used for showing/hiding the annotations on the mesh;
- Generic mesh: has all the options of the geometric template, with the addition of the "Adapt template to object" option, which starts the fitting procedure (see section 3.4).

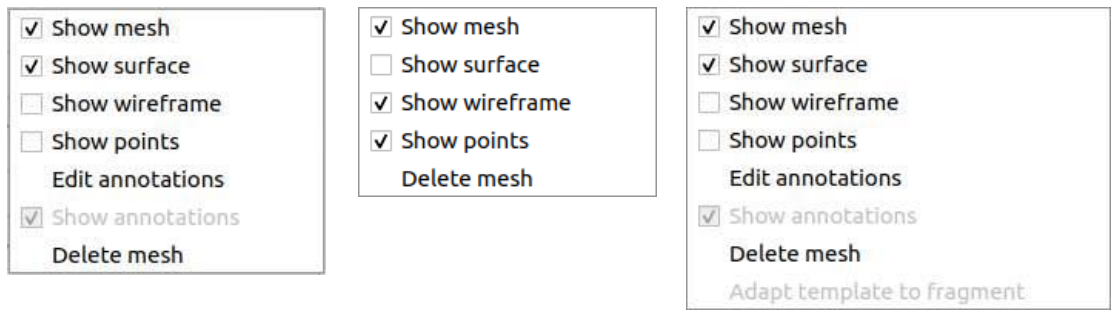

*Figure 4: on the left, the contextual menu corresponding to the right click on the template*

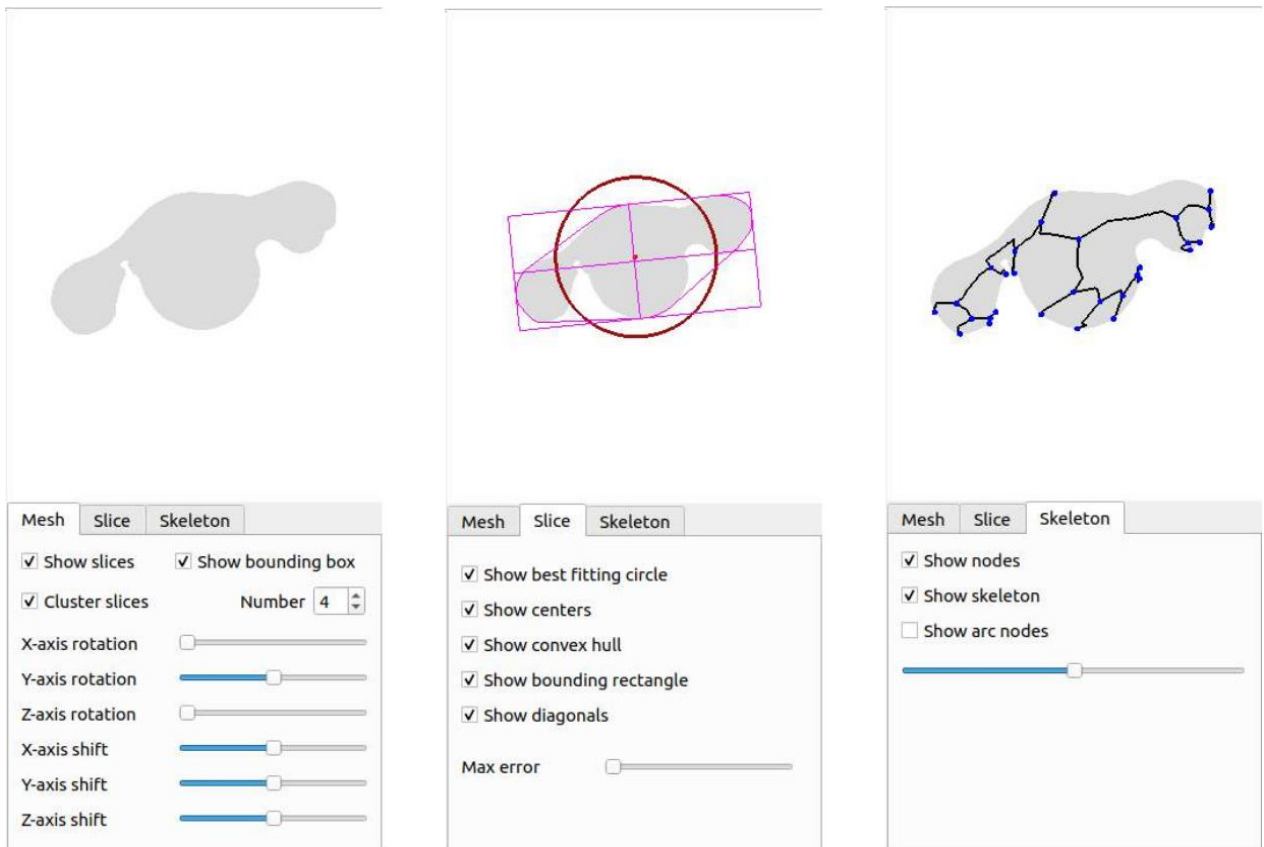

*Figure 5: the three sheets in the Slice view.*

The Slice view allows to extract information of several kinds from the template shape. It is equipped with three different sheets (see Figure 5) enclosing tools for setting and moving the plane which is used to slice the shape (sheet "Mesh"), tools for computing some shape descriptors of a slice (sheet "Slice") and computing and visualising the associated Medial Axis approximation at different levels of detail (sheet "Skeleton"). The slice under analysis is shown both in the main canvas (over the geometric template's surface) and in a smaller canvas in the upper part of the Slice view.

The Constraint view presents a drop down list for selecting the high-level constraint whose information the user is interested in. Then, the frame below is populated with the information associated to the constraint, e.g., id, type, etc. (most of information is constraint specific). This view is particularly useful for visualising the error with respect to a certain constraint. Indeed, the constrained deformation is expressed as an optimisation problem (see 1).

It follows that, in certain conditions, a constraint could not be satisfied, or several constraints could not be satisfied at the same time. Anyway, the ShapeOp library [9] will minimise the overall error, but the user must be aware of which constraints could not be fully satisfied and at which extent. The user may then take an informed decision about whether to accept, reject or change the deformation, e.g., when performing reassembly tasks.

After loading the geometric template and the cage (or computing it with the provided functionality), the "Link cage and template" action in the Edit menu is enabled (this action will compute the GBC) as well as the "Load coordinates" action in the File menu. After the computation/loading of the GBC (it may require some time, depending on the resolution of the template mesh and of the cage), the toolbar is enabled, together with its buttons, which provide the functionalities for selecting cage vertices and using them as handles to define a deformation:

The Camera button, when checked, allows only to change the view around the shape, without any other kind of interaction;

The Visible Selection button, when checked, enables a filter which allows selecting only of the vertices that are visible from the user point of view (by default it is checked);

The Vertex Selection button, when checked, allows to select vertices of the cage, either individually, or all those enclosed in a rectangle drawn by the user (the rectangle selection is performed by holding the Ctrl button on the keyboard and simultaneously left clicking on the canvas, dragging and releasing the left mouse button, see Figure 7), or all those associated to an annotation (this is obtained by holding the Ctrl button and pressing the middle button of the mouse on the annotation of interest and employs the technique defined in [10]);

The Vertex De-selection button, when checked, allows to perform the opposite operation of the previous button;

The Move Vertices button, when checked, allows to rigidly move the selected vertices of the cage. There are two possibilities: to drag (translate) the selected vertices following the mouse position (Ctrl + right button of the mouse) or to rotate them around the axis orthogonal both to the camera direction and the mouse movement1 of an amount defined by the magnitude of the mouse movement (Ctrl + left button of the mouse). The template's shape is updated at a fixed frame rate;

The Stretch Vertices button, when checked, allows to move the selected vertices away from each other in a user defined direction. The operation is achieved by holding the Ctrl button on the keyboard, left clicking on the canvas and dragging in a certain direction: the stretch direction is computed using the current and starting positions, while the amount of stretch is constant and added every frame. As in the previous point, the template's shape is updated at a fixed frame rate.

Finally, the Main window is equipped with a footer where the number of vertices, edges and triangles of the template and of the cage are shown.

## 3.2 Annotation window

When the user clicks on the "Edit annotation" voice in the contextual menu of any mesh (cage excluded) a new window, called Annotation, is opened. Indeed, the user can either annotate the geometric template, thus actually adding the link with the semantic template, defined elsewhere as a knowledge formalisation or based on her/his own contextual knowledge, or annotate a specific object of the class for documentation purposes. It recalls the general schema of the Main window, with a menu bar, a toolbar, a sidebar and a central canvas where meshes and other entities are drawn.

Here, the user can select mesh vertices and add tags to create annotations, take measurements and add attributes to annotations, set relationships among annotations and set constraints on attributes and relationships.

In the current version, the menu bar contains the File menu only: it provides functionalities for saving, loading and clearing (deleting all) the annotations, clearing the selections and saving/loading the constraints/relationships.

The toolbar contains several buttons:

The Camera button, when checked, allows only to change the view around the shape, without any other kind of interaction;

The Restore Camera button, when pressed, resets the camera to the original displacement;

The Visible Selection button, when checked, enables a filter which allows the selection only of the vertices entities which are visible from the user point of view, depending on the current interaction modality (see next points - by default is checked);

The Eraser button, when checked, changes to the de-selection modality for discarding the newly selected entities (vertices, edges, triangles, annotations) depending on the interaction modality (see next points);

The Points Rectangle button, when checked, changes the selection modality to points selection, i.e., allows to select (or de-select if the Eraser button is checked) vertices of the mesh, either individually or all the ones included in a rectangle drawn by the user (following exactly the same procedure of the Vertex Selection button in the previous sub-section, see Figure 7);

The Polylines button, when checked, changes the selection modality to edges selection, i.e., allows to select (or de-select if the Eraser button is checked) edges of the mesh, by picking some successive points, in an additive way, on the surface (see Figure 7): the path between them is automatically computed by the system following an approximation of the Dijkstra algorithm (the search for the shortest path is interrupted when the target vertex is found);

The Triangles Rectangle button, when checked, changes the selection modality to triangles selection, i.e., allows to select (or de-select if the Eraser button is checked) triangles on the surface of the mesh, either individually or all the ones included in a rectangle drawn by the user (see Figure 7);

The Lasso button, when checked, changes the selection modality to triangles selection, i.e., allows to select (or de-select if the Eraser button is checked) triangles on the surface of the mesh following the common lasso metaphor [Las20], i.e., the user draws a "polygon" on the surface picking successive points (as for the Polyline button, but now the polyline must be closed), then right-clicks in the interior of the "polygon" (this is necessary, since the interior of a polygon over a 3D surface is not well-defined) and the triangles enclosed in it are automatically selected);

The Annotate button, when clicked, opens a dialog asking information about the annotation to be saved;

The Edit button, when clicked, allows to modify the selection associated to an annotation (it requires that one and only one annotation is selected). The operation can be completed re-annotating the selection (thus using the Annotate button);

The Annotation Selection button, when checked, changes the selection modality to annotation selection, i.e., allows to select annotations from the mesh surface. This can be obtained by holding the Ctrl key and left clicking over the annotation: if more than one annotation is under the mouse when the user clicks the left mouse over the surface, a dialog is opened to ask the user which annotation she/he wants to select;

The Add Constraint button, when clicked, displays a dialog to input the type and other properties of the constraint, based on the selected annotations. Generally speaking, the dialog (see Figure 10) allows to insert the weight associated to the constraint, the minimum and maximum value of a range (in the constraints where it makes sense). This button is shown only if the current mesh the template;

The Relationships Graph Construction button, when clicked, performs the extraction of some basic relationships among the annotations, derived by geometric analysis. Currently, the containment and adjacency relations are extracted. This button is shown only if the current mesh is that of the template;

The Show Relationships button, when clicked, opens a window (see next sub-section) showing the relationship graph. This button is shown only if the current mesh is that of the template;

The Constrain button, when clicked, initialises the ShapeOp library to act on the constraints defined until now (this typically requires some time, depending on the resolution of the template shape and of the cage and on the number of defined constraints). This button can be found only if the current mesh is that of the template. This is disabled until a cage is loaded and the BC are computed/loaded;

The Withdraw Constraints button, when clicked, clears the ShapeOp library status, thus reverting to the state of the system before the Constrain button was clicked. This button is enabled only after the Constrain button is clicked;

The Transfer button is used to transfer the annotations from one source mesh to a target one. When clicked, it opens a dialog for the selection of a mesh file and then starts the transfer procedure (see [5]). Currently, this switch the new mesh with the original one in the system.

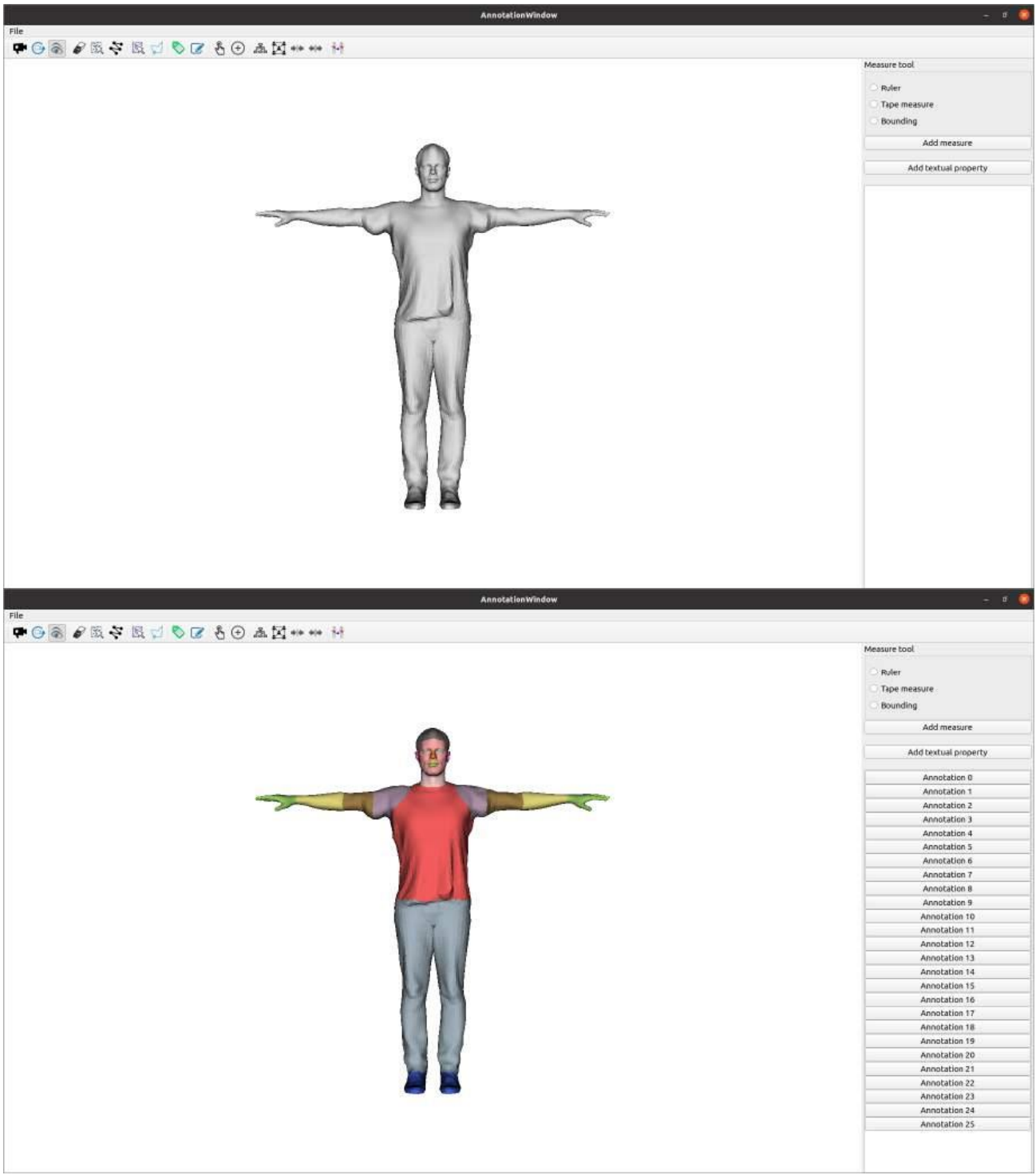

*Figure 6: the Annotation window with (bottom) and without (top) annotations.*

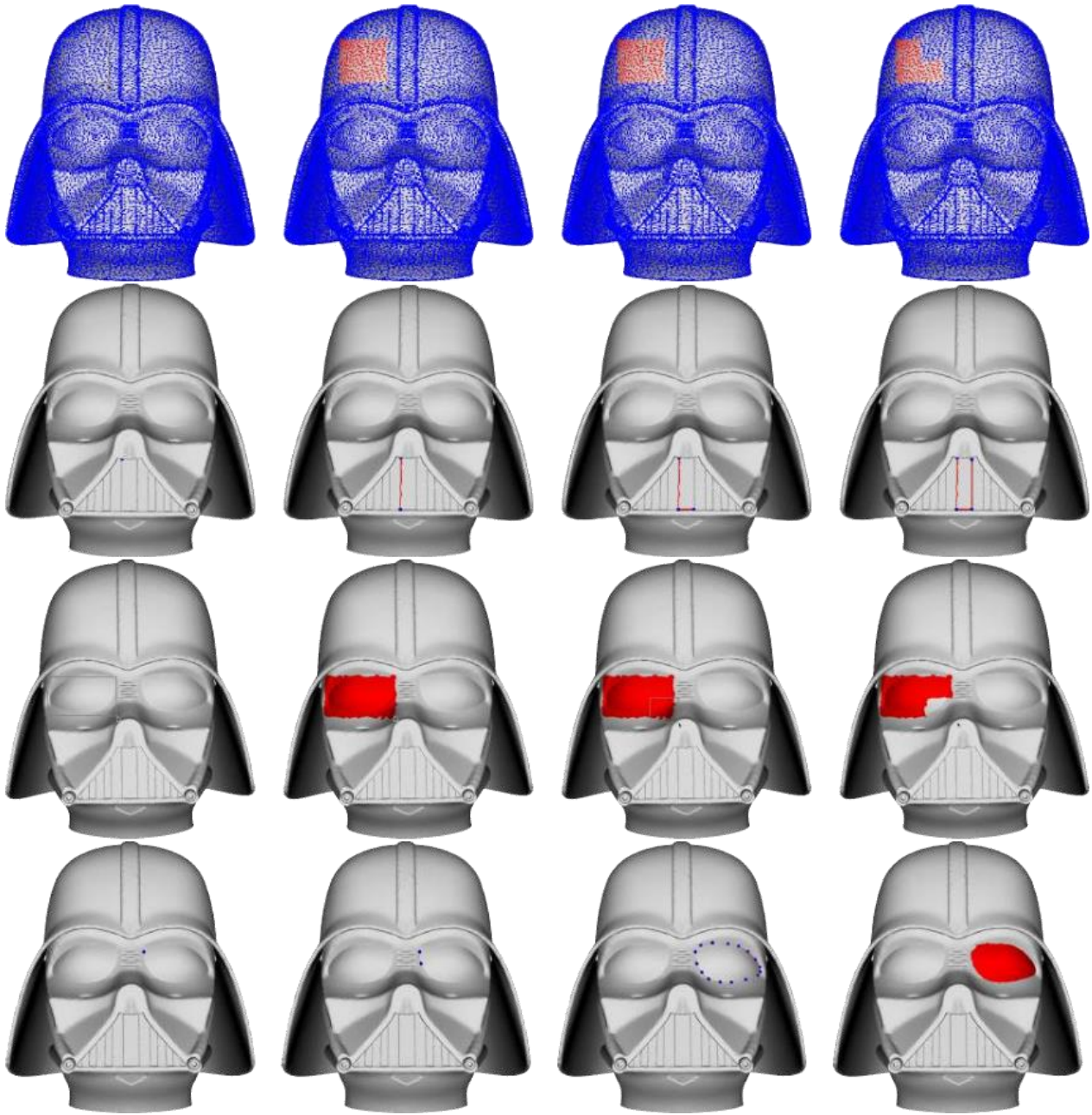

*Figure 7: Different types of selections: going downwards selection of points, edges and triangles, there the last two rows depict two different tools for the triangles selection, namely rectangle and lasso selection.*

The sidebar can be viewed as the composition of two different parts:
- In the upper part, the user finds the tools for adding attributes to annotations. To do this, the user has first to select one annotation (an error message is displayed if the user tries to add an attribute to no or more than one annotation at a time). This software considers two kinds of attributes: qualitative and quantitative. To add a qualitative attribute, the user has to press the "Add textual property" button, opening a dialog asking for a name of the attribute and some free text (it can be used for taking notes, adding meta-data such as the place of finding of a piece. etc.). In the current version of the software, only the possibility to add measures as quantitative properties is provided, while other more complex geometric properties (e.g., mean curvature, shape index, ecc.) will be integrated in the future, and in particular they can be taken using three kinds of tools (see Figure 8):
  - Ruler: it requires to select two points over the surface (Ctrl+left click, it works only on the selected annotation) for defining the extrema of the measure (Euclidean distance);
  - Tape measure: it works similarly to the Polylines selection, i.e., it requires to select points on the selected annotation (Ctrl+left click) in an additive way, while computing the shortest path between the couples of successive points to define the geodesic measure (approximated);

- ○ Bounding measure: this tool is useful for defining measures between extrema not falling on the surface. It requires to define a direction for the measure (Ctrl+left click and drag) and returns the distance between the two farthest points of the annotation in that direction. To compute this measure, points are projected on a line having the direction set by the user and passing through the barycentre of the annotation. Two clipping planes are shown when defining the measure to help the user visualise the measure.
- In the lower part, a list of annotations of the mesh is shown, were each voice can be expanded for showing some properties, such as id and tag, and the list of associated attributes, that in their turn can be expanded to show the associated properties, a button for removing the attribute and, in the case of a measure, a checkable button for showing/hiding it in the canvas (see Figure 9).

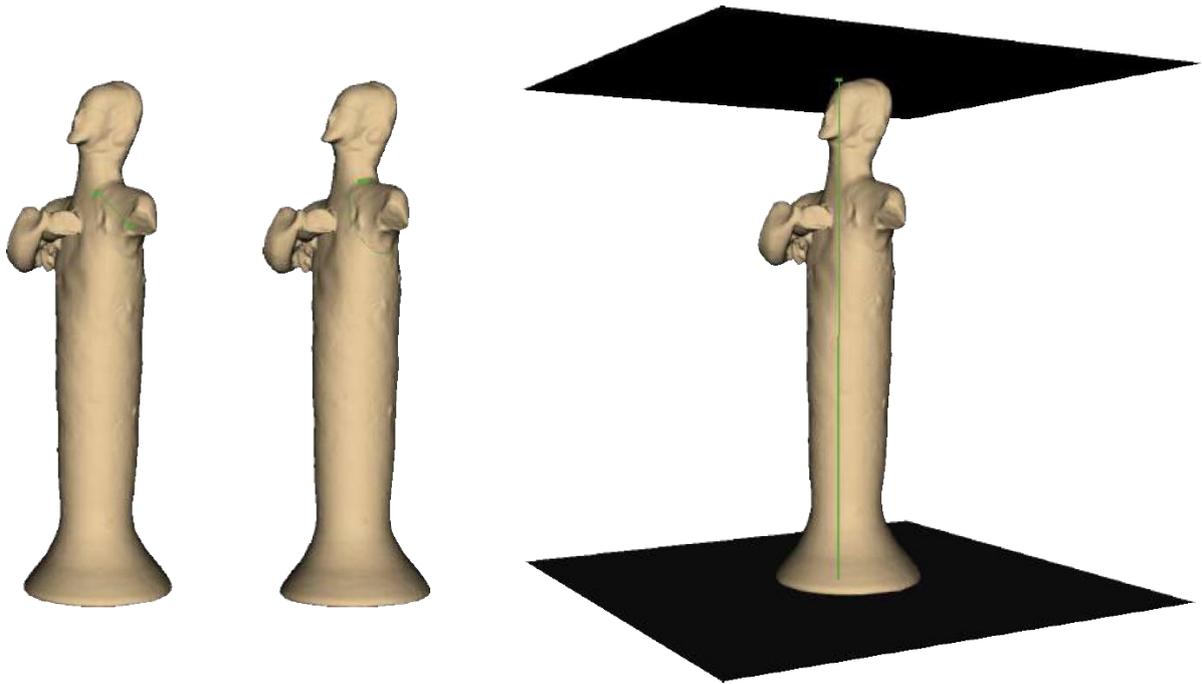

Figure 8: from left to right, examples of measures taken with the ruler, tape and bounding tools.

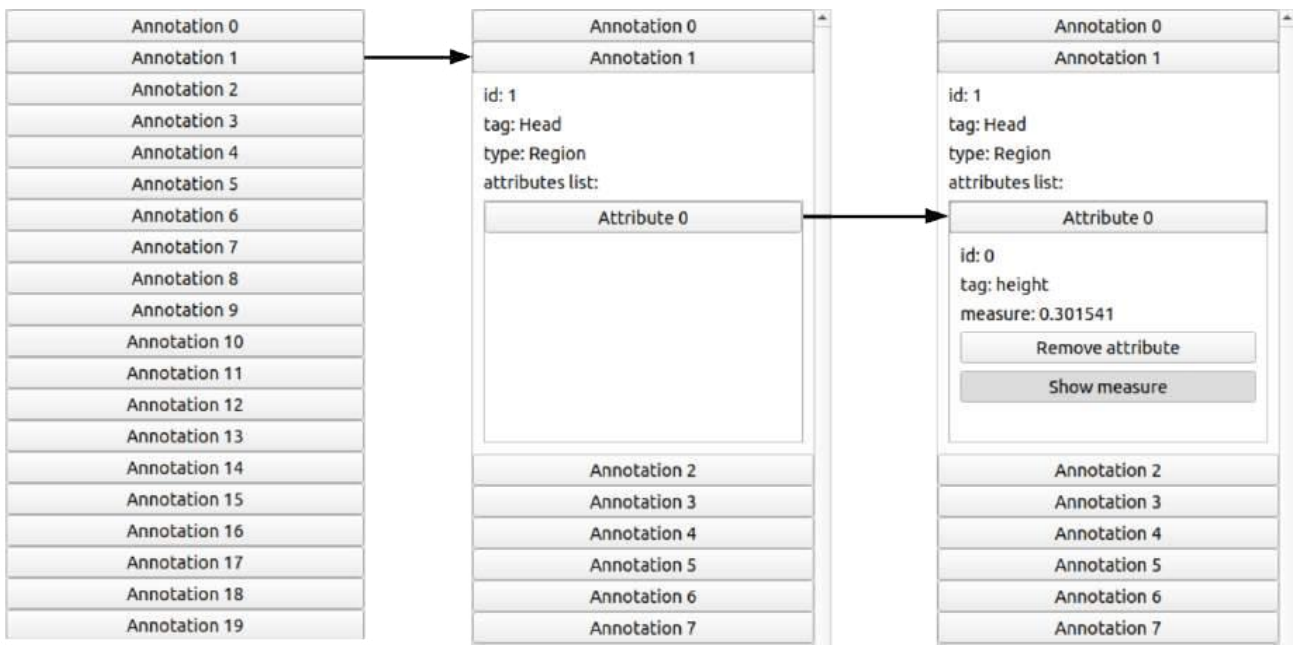

Figure 9: the simple list of annotations (left), the list with an expanded annotation (middle), the list with an expanded annotation with one of its attributes expanded (right).

## 3.3 Relationships window

In this window (see Figure 11), the system provides an overview of the stated relationships between annotated parts. The relationships are represented as line segments or arrows, depending on the direction of the relationship, between nodes of a graph, representing annotations. This window also allows the creation of new relationships. The user can select two or more annotations (simply left-clicking on any node in the graph) and then press the button, which calls a dialog for selecting the type and specifying the properties of the new relationship. Moreover, the user can also constrain the existing relationships, by pressing the button) and so initialising the ShapeOp library (same behaviour as in the Annotation window).

When passing the cursor over an annotation (node of the graph) or a relationship, the sidebar displays related information regarding the specific entity (see Figure 12).

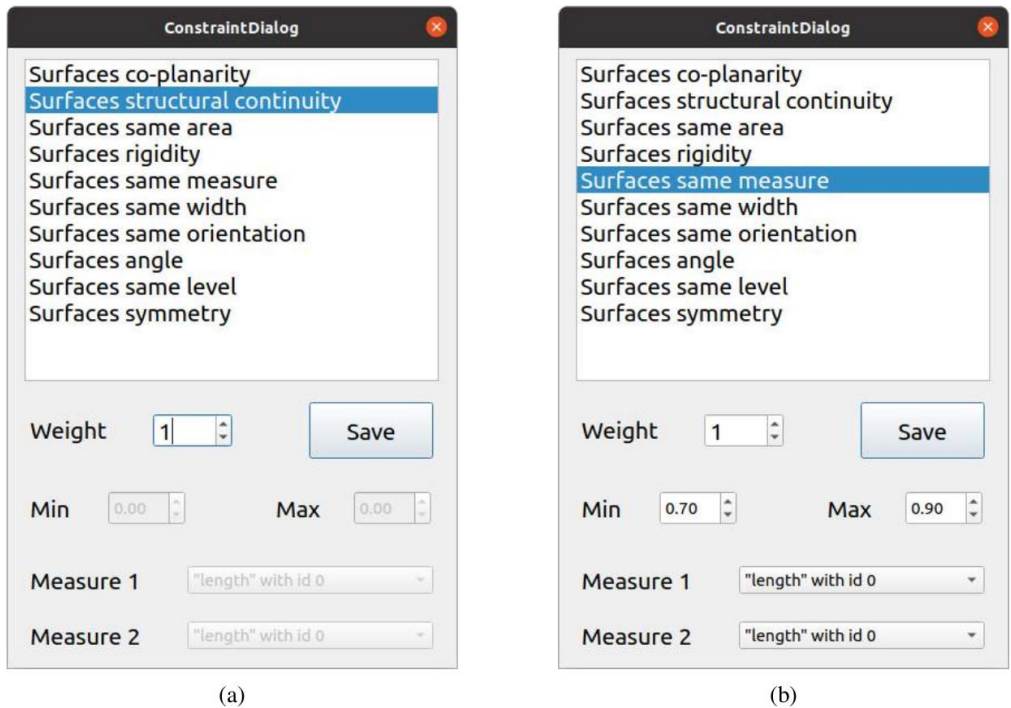

(a)                                           (b)

Figure 10: the constraint dialog enables the input fields depending on the type of ship/constraint. In the current implementation the measures are constrainable only if the relationship is between 2 annotations.

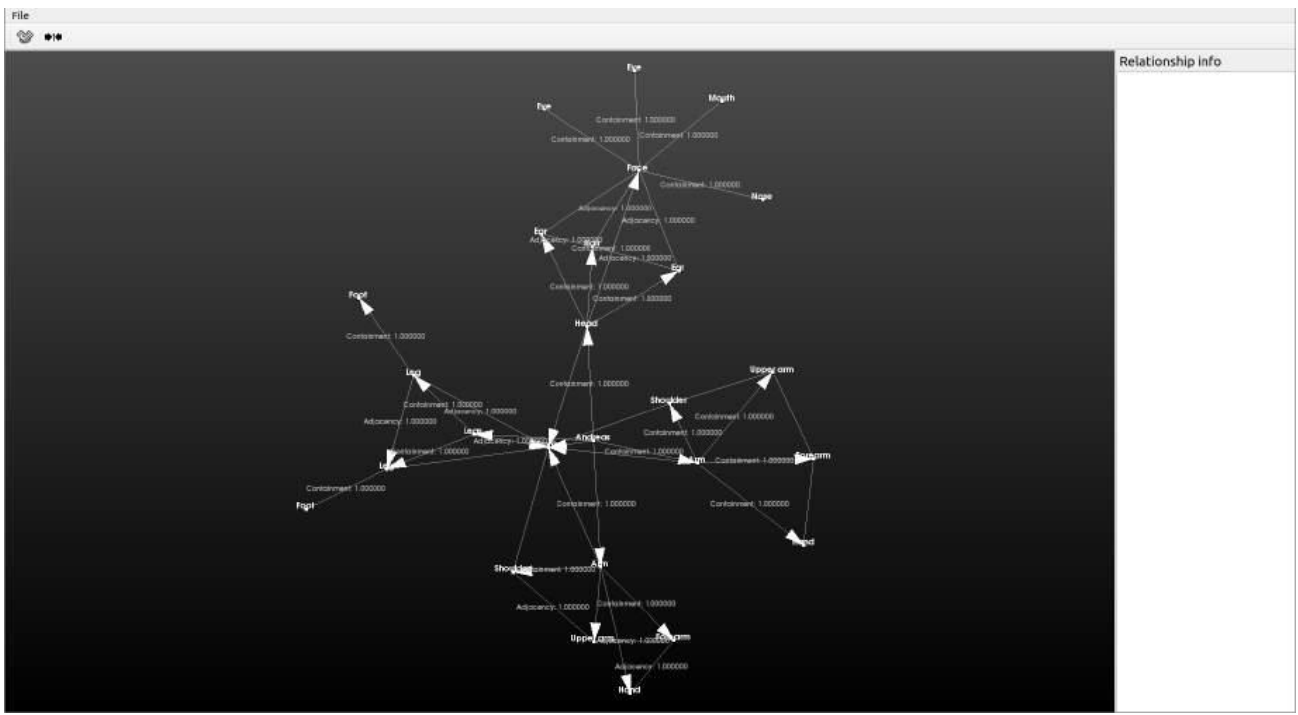

Figure 11: the Relationships window

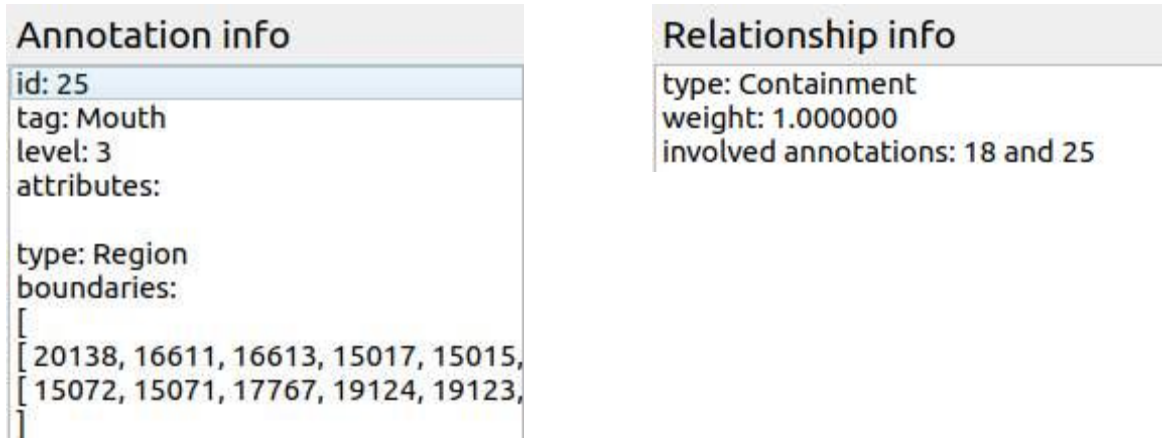

*Figure 12: different information shown when passing over annotations (left) and relationships (right) with the cursor.*

## 3.4 Fitting procedure

After clicking on the "Adapt template to object" option, described in section 3.1, the fitting procedure takes place. First of all, the system checks if at least three *landmarks* (i.e, annotated points) are in common between the selected fragment and the parametric template. If otherwise, the system requires the user to manually select three common points on the shape of both using a dedicated dialog (see Figure 13).

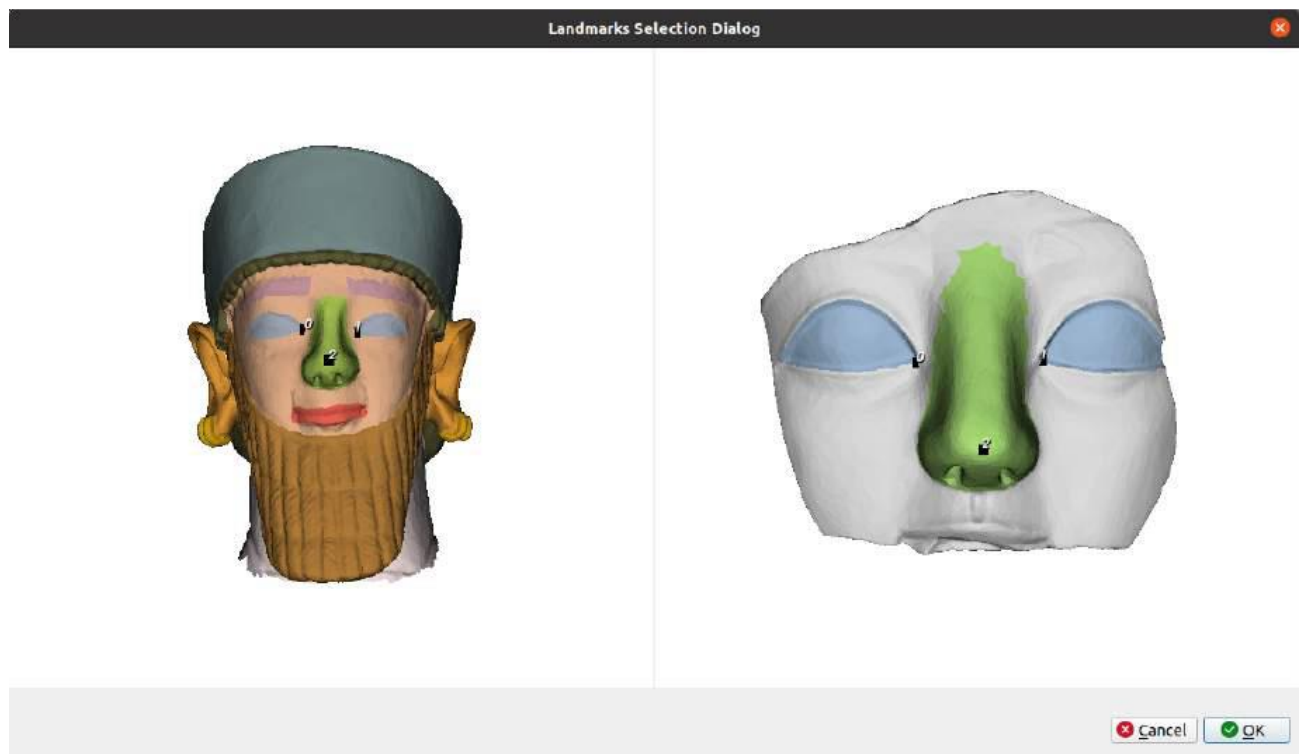

*Figure 13: the dialog window for inserting the landmarks for the rigid alignment step.*

The system automatically scale the template to fit the size of the fragment and translates/orients the fragment to place it on the correct position of the template (rigid alignment, made exploiting Umeyama's algorithm [11]). This allows to perform a rough alignment between the fragments and the template, in a least-square manner. The results of this step can be seen in Figure 14.

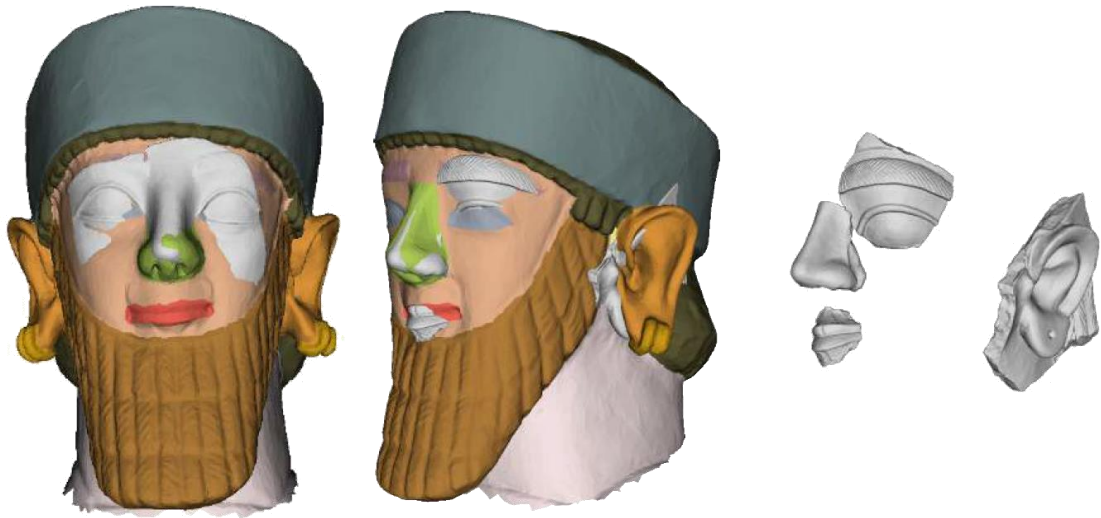

*Figure 14: the results of the rough alignment obtained employing Umeyama's algorithm.*

Next, the system deforms the shape of the template in a non-rigid manner so that it fits more precisely the shape of the fragment. This is defined as a combined optimization problem taking into account two separate problems: the definition of *correspondences* between the template and the fragment and th deformation of the template to fit the fragment while preserving the defined constraints (details in [1]). The results with the current version of the software can be seen in Figure 15.

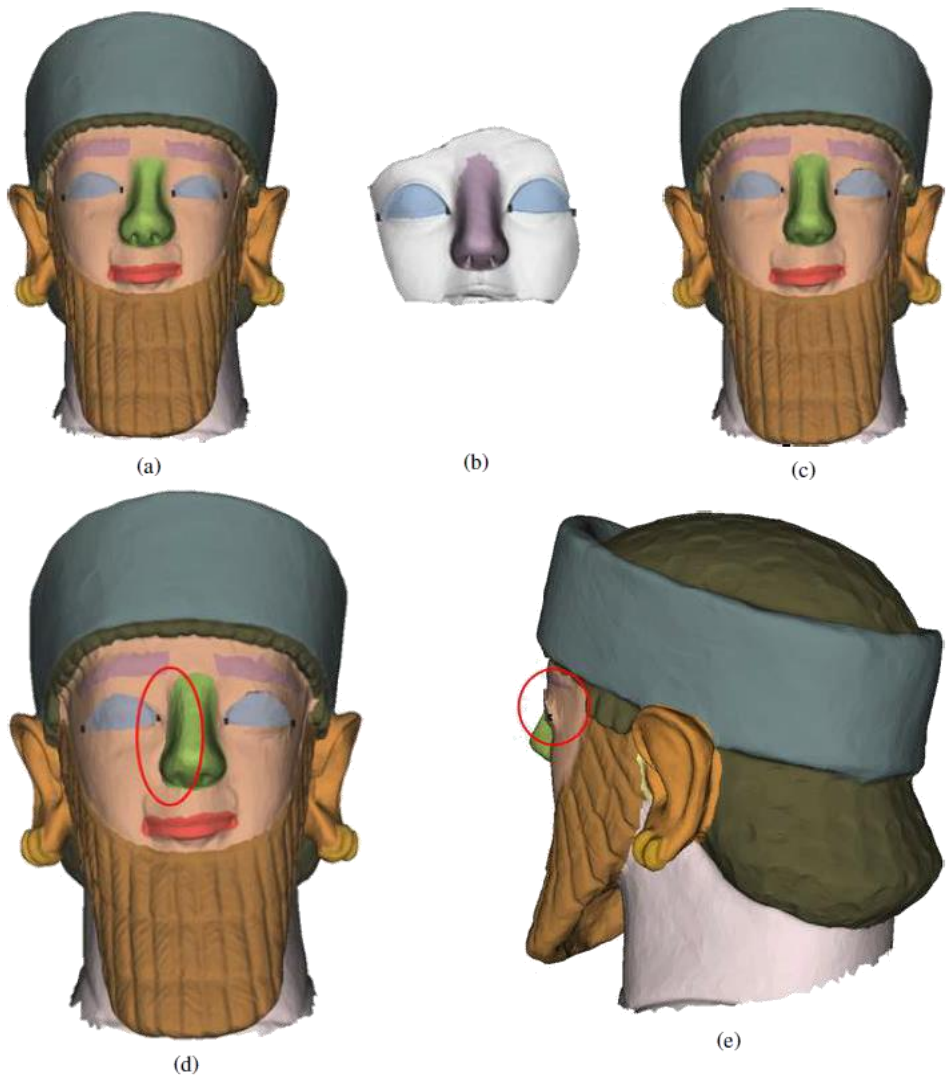

*Figure 15: the first result in the non rigid fitting step: (a) template, (b) annotated fragment, (c) deformed template, (d) first distortion highlighted, (e) second distortion highlighted.*

## 4. File formats

The formalisation of the semantics of a certain class of objects is (at least currently) a long work which can be performed by one or more experts. For this reason, tools should be provided to store the results produced so far in such a way that could be easily shared between the experts and provided to the final user.

In this software, such results are stored exploiting the JavaScript Object Notation (JSON)© [12] format, which is hugely employed in several fields thanks to its simplicity. In particular, here it is used for storing annotations and related info (i.e., tag, type, attributes, etc.) and relationships/constraints (better, the relationships' graph). The library employed for the writing/parsing of the files is RapidJSON© [13].

Other possible results are given by the deformation of the geometry of the template: this can be saved, as well as the employed cage, using any of the standard file formats for triangular meshes (e.g., .stl, .ply, .obj, .off, etc.). Finally, the generated GBC can be saved as a matrix, where the rows correspond to the coordinates referred to the template vertices and the columns correspond to the coordinates of the cage vertices. In this software, they are saved as textual files (.txt) where the first row reports the generation procedure (for instance "Mean Value Coordinates") and the second row reports the number of vertices of the template and of the cage (useful for low-level IO). Remaining rows contain the GBC matrix.

Notice that if "Green Coordinates" is the employed method, the coordinates corresponding to triangles are stored in the tail of the common ones (see GC in [4]).

### 4.1 Annotation file format

The annotation file format contains as root the object annotations, which is a list of annotation objects containing the following fields:

- id: the integer number identifying the annotation in the system;
- tag: a string (in future extensions this field will be substituted with a more generic information field containing a reference to an object, being it a string, an image, or whatever) associated to the annotation;
- colour: a triplet (more formally, a list of 3 integers) defining a colour associated to the annotation (in the future this field will be removed and the colour will be automatically set by the application GUIs);
- attributes: a list of attribute objects specified in the following;
- type: the type of selector of the annotation. Currently there are 3 possible values: "point", "line" and "region".
- Depending on the value of the type field, this field can have the following values:
  - points: only when the type field has "point" value. It is a list of indices of mesh vertices;
  - polylines: only when the type field has "line" value. It is a list of lists of indices to successive vertices of the mesh, defining one or more poly-lines;
  - boundaries: only when the type field has "region" value. It is a list of lists of indices
  - to successive vertices of the mesh, enclosing bounded patches on the mesh surface

The attribute objects contain the following fields:

- id: the integer number identifying the attribute within the annotation;
- name: a string containing the name associated to the attribute (e.g., "height");
- type: a string defining the type of the attribute (currently only "semantic" and "measure");
- there are two possibilities for the final part of the object, depending on the value of the previous field:
  - note: only if the value of type is "semantic". It is a string containing a free text.
  - measure: an object containing the following fields:

- **tool**: a string specifying the tool used for taking the measure (currently only "ruler", "tape" and "bounding");
- **points**: a list of indices to the mesh vertices involved in the measure. Depending on the previous field, they can be 2 or more (2 for an Euclidean distance, more for the approximate geodesic one).
- **direction**: this field is present only if the previous value is "bounding". It is a vector (more formally a list of 3 doubles) defining the measure direction.

## 4.2 Graph file format

The graph file format contains as root the object's relationships, which is a list of relationships objects containing following the same base structure:

- **id**: an integer number identifying the relationship;
- **type**: a string identifying the relationship. There are several possible values, one for each kind of constraint present in ShapeOp (the framework always gives the possibility to use the geometric constraints) plus the high-level constraint defined in [1]. Notice that not all the relationships are constraints, so here possible values are even "containment" and "adjacency";
- **isDirected**: a boolean value for understanding if the arc correponding to the relationship is directed or not;
- **annotations**: a list of indices to the annotations involved in the relationships;
- **isConstraint**: a boolean value stating if the relationship is a constraint;
- **weight**: a double value associating a weight to the constraint. This field is present only if the previous is true;
- **constraint**: is an object reporting all the parameters of the constraint (so this field is present only if isConstraint is true). The inner structure of this object is really dependent on the constraint, e.g., the "Proportion" constraint (see [1]) requires measure1 and measure2 indices identifying the measure to be constrained both on the first and the second annotation and a minValue and maxValue defining an acceptance range.

The recovery of the graph is simply achieved by creating a node for each annotation and then scrolling the relationships list and creating arcs accordingly.

# 5. Employed technologies & libraries

| Technologies & libraries | Version | Usage |
|---|---|---|
| **CMake** | 2.8.8 | Mangement, compilationa nd linkage of the whole project |
| **C++** | 11 | Development of the whole software (libraries excluded) |
| **Eigen** | 3.3.9 | All the linear algebra operations |
| **Git** | 2.28 | Version control and distribution of the source control |
| **Google OR-Tools** | 8.2 | Management, definition and solution of the optimization problems related to the fragments fitting |
| **ImatiSTL** | 4.1 | Management of triangular meshes |
| **JSON** | - | File format for storing annotations and relationships |
| **MathGeoLib** | 1.5 | Extraction of some geometric properties of shapes (e.g., Oriented Bounding Box) |
| **Nanoflann** | 1.3.2 | Management of KD-trees for performance improvement |
| **Qt** | 5.12.9 | Design and development of the GUI and event management |
| **RapidJSON** | 1.1.0 | Management of JSON files |
| **ShapeOp** | 0.1.0 | Constraints definition and solution of the related optimization |
| **Shewchuk's Triangle** | 1.6 | Triangulation of the polygons defining shape slices. Indeed, the computation of some slices' properties (e.g., approximate medial axis) requires a triangulation of the plane to work. |
| **Visualization Toolkit (VTK)** | 7.1.1 | Visualization of all the geometric entities of the software, from the meshes to the slices, passing through annotations and graphs |